# USE OF EPICS FOR HIGH-LEVEL CONTROL OF SNS CONVENTIONAL FACILITIES [1]

J.K.Munro, Jr., J.E.Cleaves, E.L.Williams, Jr., D.J.Nypaver, ORNL, Oak Ridge, TN 37831, USA
K.-U.Kasemir, LANL, Los Alamos, NM 87545, USA
R.D.Meyer, Sverdrup Technologies, Tullahoma, TN 37388, USA

Abstract

The SNS Project intends to integrate Conventional Facility Controls with its EPICS-based Accelerator and Target Control Systems. EPICS will therefore be used to provide distributed high-level access to all subsystems of the SNS conventional facilities, including cooling water towers, chilled water, de-ionized water, HVAC, waste processing, and power monitoring. EPICS will provide operator displays and high-level control for more than 1000 process variables. EPICS support will be provided by four IOCs using PowerPC-based VMEbus controllers. Low-level control will be provided by Allen-Bradley ControlLogix PLCs that will communicate among themselves using ControlNet and with EPICS using EtherNet/IP [2]. Both the PLC layer and the EPICS layer will be implemented by an industrial supplier. File server support will be Linux-based and Concurrent Versions System (CVS) will be used to manage version control, both for EPICS and for PLC program and configuration files. All system process variable names, hardware, software, and database configuration properties will be maintained in a master Oracle database which will be used to generate and maintain EPICS and PLC databases for the entire project.

## 1 INTRODUCTION

This paper describes the business and technical arrangements made by the SNS Project with Sverdrup Technologies (SvT) to develop the control systems for the SNS conventional facilities (CF). EPICS will be used to provide high-level control and display of information about CF subsystems to the operators; commercial PLC technology will provide low-level control of all CF hardware. Primary CF subsystems include HVAC, chilled water, de-ionized water, air handling, waste processing, and power monitoring.

New elements for this project exist in both the technical and business areas. In the technical area the Linux operating system is being used to support all EPICS functions and development tools; the Motorola MVME2101 PowerPC processor board is being used in all I/O controllers (IOCs). In the business area this development effort probably is the first time all CF control systems are being integrated "up front" into the EPICS control system for a large accelerator facility. This infrastructure must be built early to provide support for the major accelerator subsystems as they are added; hence, is one of the first major SNS subsystems to be completed and put into operation. To guide development, standards and guidelines for control system software and operator displays have been prepared, approved, and distributed to the development contractor(s). As a consequence, the CF control system development schedule has driven the development of standards and guidelines and selection of hardware that will be used for development of control systems for other SNS subsystems.

The basic plan is for the SvT developers to identify all the I/O points required for the CF. I/O point names (following a standard naming convention) and descriptions are entered into spreadsheets from which they can be read by scripts to generate tag name lists used in the PLCs and process variable (PV) names for EPICS databases which are loaded into IOCs. Tag names are grouped by EPICS record type and packed into data transfer arrays which are used by a communications driver (EtherNet/IP) [2] to move data for the operator displays between the PLCs and IOCs. The SvT developers will write the ladder logic programs used by the PLCs for low-level control of the hardware, will develop the operator screens needed for high-level control, and will generate the EPICS databases to be loaded into the IOCs.

## 2 BUSINESS ARRANGEMENTS

All SNS CF control systems are being developed through fixed price contracts with SvT, located in Tullahoma, TN, a few hours' drive from the SNS Project office in Oak Ridge. SvT engineers have broad experience in the integration of highly technical systems. Contracts have been placed for system (Title I) design, detailed (Title II) design, cabinet fabrication design, electronics procurement and cabinet fabrication, software and database development (both EPICS and PLC logic), loop testing after equipment installation, and procurement and storage of sensors and control elements. Cabinet, sensor, control element, cabling/conduit installation, and field cabling termination will be done via fixed price contract with an installation contractor with expertise in this area.

Having a single integrating contractor perform all this work eliminates many interfaces and the need to pass information from one contractor to another. It also allows loading developed software into cabinet equipment and testing the entire system before delivery for installation.

## 3 PLC APPLICATION DEVELOPMENT ENVIRONMENT

All PLC application development is being done using RS Logix 5000 on a Windows 2000 system to configure the hardware and write the ladder logic control programs. All I/O point information is being maintained in Excel spreadsheets and MS Access databases.

Considerable effort went into the development of a standard naming convention for control PVs. The results of this effort have been published in a project reference document [3]. Scripts have been written to use the common signal naming syntax to read names from the I/O points database maintained in Excel spreadsheets and write files for the PLC and EPICS databases with PV names constructed according to the different syntax rules used for tag names (PLCs) and the corresponding EPICS database records.

Performance measurements made for data transfers between PLCs and IOCs using the EtherNet/IP driver [2] showed in dramatic fashion the importance of using arrays to transfer data quickly and efficiently. For this reason a separate data transfer array is used for each different EPICS record type, e.g., ai, ao, bi, bo, mbbi, and mbbo.

## 4 EPICS APPLICATION DEVELOPMENT ENVIRONMENT

The computers and software used by SvT were first set up in Oak Ridge the way they would be configured in Tullahoma for the development work. All software tools were installed and configured. An example application was set up and used to check out every step of the development sequence, from I/O points listed in an Excel spreadsheet to an operating application on a PLC with data transferred to an EPICS database loaded into an IOC using the EtherNet/IP driver, and PVs displayed on screens built using the Extensible Display Manager (EDM). This example application was later used in training exercises.

A minimal development environment sufficient to duplicate the environment set up at SvT was set up in Oak Ridge for use in providing technical support to the SvT team. This system is also available for giving demonstrations at SNS program review meetings. Any upgrades to operating systems, application software, or development procedures are first implemented and tested on the Oak Ridge environment before being installed on the Tullahoma system. This minimizes impacts on the developers and supports strong configuration control.

### 2.1 Hardware Configuration

Seven workstation class (Pentium III Xeon) computer were purchased for programming the PLCs (4 Windows 2000 systems) and developing the EPICS screens and databases (3 Linux systems). Two Linux file servers with 5 disk RAID systems were purchased. Identical EPICS application development environments (ADEs were set up on them. A second Ethernet card was installed in the SvT server to permit setting up the development environment at Tullahoma on a private LAN. These two servers communicate via a T1 line.

### 2.2 Software Configuration

An SNS CF ADE was set up to minimize the amount of EPICS system administration work needed by SvT. The directory structure is consistent with conventions used by the scripts, rules, and makefiles provided with the official EPICS software distribution and described in documents available from the Advanced Photon Source Web site at Argonne National Laboratory.

### 2.2 Application Development Process

The SvT development team has already identified the CF I/O points from process and instrument drawings (P&IDs) and has used spreadsheets and an MS Access database to archive and maintain this data. Scripts were written to apply the tag and EPICS PV naming conventions to this I/O point data to generate a set of spreadsheets which are used as input to RS Logix 5000 to generate tag names and to scripts used to build the EPICS database files needed for the operator displays. The tag names are used in the ladder logic programs which are downloaded into the PLCs. Data for selected tag names are packed into transfer arrays which an EtherNet/IP driver reads from PLC memory and writes into the EPICS database records accessed by the displays.

All these sets of I/O points data will eventually be loaded into the SNS master Oracle database and maintained by tools written for this purpose. A long-term goal is to automate the process of generating and maintaining the data that must be downloaded to the PLCs and IOCs from the master Oracle database so that all control system data can be centrally maintained and administratively controlled. This should also allow IOC databases to be modified and downloaded more reliably and IOCs to be rebooted automatically.

Training of the development team was built around the development sequence and processes that would be used to generate all parts of the CF control system. It consisted of an abbreviated version of the standard EPICS course, with emphasis on the more limited tool set needed to build the CF applications.

## 5 CONFIGURATION CONTROL

During development the SvT team will arrange for configuration control of the I/O points database, scripts, configuration files, PLC programs, and operator display screens. All primary data, scripts, programs, and display screen files from SvT will eventually be archived and maintained in SNS project CVS repositories.

Configuration control of all EPICS related software is done using the SNS CVS repository in Oak Ridge. New versions of the screen editor and display manager EDM are used and tested in Oak Ridge before being installed on the file server in Tullahoma. The same procedure will be followed for any changes to EPICS base or any of the EPICS extensions obtained from the primary repository at Argonne National Laboratory.

## 6 CONCLUSIONS

A number of lessons have been learned in the areas of developing requirements, functional system design documents, standards, training, software development environment, technical support, configuration management, and contract management from experience to date. This effort is leading the way for a significant amount of control system application development to follow for SNS. A few highlights will be indicated here.

Document tag names, signal names, setpoints, EPICS record parameters, and logic functions in a manner that can be used to test completed software (e.g., in spreadsheet or database applications). Generate as much of this information as possible from P&ID diagrams: this permits converting design information in P&ID language to data in formats that can be used directly and tested by software applications. The contractor who will develop the CF control systems should generate all the design information before any control system development activities begin: this should help to insure that the contractor software developers understand what must be built.

Standards should encompass all aspects of the work, should be documented in a design criteria or similar document, then applied to some representative prototypical examples and tested to verify their features accomplish what is desired/expected.

Choose examples for use in training that can be used later in what must be developed. These examples must use the standards and tools in the same way as the product to be delivered. Set the examples up in the same ADE (in a "training" directory) that will be used for development.

In preparing the ADE, make reasonable efforts to use database application and other tools with which the contractor is already familiar. This should speed development time and support development of a more reliable product.